\documentclass[twocolumn,floatfix,superscriptaddress,a4paper,showpacs,showkeys,nofootinbib,reprint,prc,noeprint]{revtex4-1}
\usepackage[english]{babel}
\usepackage[utf8]{inputenc}

\usepackage[colorlinks=true,linktocpage=true,linkcolor=blue,citecolor=blue,allcolors=blue]{hyperref}
\usepackage{url}

\usepackage{epsfig}
\usepackage{float}

\usepackage{amsmath, amssymb, physics} 

\usepackage{enumerate, color, xcolor, latexsym, xfrac, needspace}

\begin{document}

 \title{Resonance suppression during the hadronic \\ stage from the FAIR to the intermediate RHIC energy regime}

\author{Amine Chabane}
\affiliation{Institut f\"{u}r Theoretische Physik, Goethe-Universit\"{a}t Frankfurt, Max-von-Laue-Str. 1, D-60438 Frankfurt am Main, Germany}
\affiliation{Helmholtz Research Academy Hesse for FAIR (HFHF), GSI Helmholtzzentrum f\"ur Schwerionenforschung GmbH, Campus Frankfurt, Max-von-Laue-Str. 12, 60438 Frankfurt am Main, Germany}

\author{Lisa Engel}
\affiliation{Institut f\"{u}r Theoretische Physik, Goethe-Universit\"{a}t Frankfurt, Max-von-Laue-Str. 1, D-60438 Frankfurt am Main, Germany}

\author{Tom Reichert}
\affiliation{Institut f\"{u}r Theoretische Physik, Goethe-Universit\"{a}t Frankfurt, Max-von-Laue-Str. 1, D-60438 Frankfurt am Main, Germany}
\affiliation{Frankfurt Institute for Advanced Studies (FIAS), Ruth-Moufang-Str. 1, D-60438 Frankfurt am Main, Germany}
\affiliation{Helmholtz Research Academy Hesse for FAIR (HFHF), GSI Helmholtzzentrum f\"ur Schwerionenforschung GmbH, Campus Frankfurt, Max-von-Laue-Str. 12, 60438 Frankfurt am Main, Germany}

\author{Jan~Steinheimer}
\affiliation{Frankfurt Institute for Advanced Studies (FIAS), Ruth-Moufang-Str. 1, D-60438 Frankfurt am Main, Germany}

\author{Marcus Bleicher}
\affiliation{Institut f\"{u}r Theoretische Physik, Goethe-Universit\"{a}t Frankfurt, Max-von-Laue-Str. 1, D-60438 Frankfurt am Main, Germany}
\affiliation{Helmholtz Research Academy Hesse for FAIR (HFHF), GSI Helmholtzzentrum f\"ur Schwerionenforschung GmbH, Campus Frankfurt, Max-von-Laue-Str. 12, 60438 Frankfurt am Main, Germany}

\date{\today}

\begin{abstract}
The energy and centrality dependence of the kaon resonance ratio $(K^{*0}+\Bar{K}^{*0})/(K^+ + K^-)$ is explored in the RHIC-BES and CBM-FAIR energy regime. To this aim, the Ultra-relativistic Quantum Molecular Dynamics (UrQMD) model is employed to simulate reconstructable $K^{*}$ resonances in Au+Au and p+p collisions from $\sqrt{s_{\text{NN}}}=3-39$ GeV. We obtain a good description of the resonance yields and mean transverse momenta over the whole investigated energy range. The decrease of the $K^*/K$ ratio, with increasing centrality is in line with the available experimental data. We also observe the experimenatlly measured increase in $\langle p_{\text{T}}\rangle$ with increasing centrality which is interpreted as a lower reconstruction probability of low-$p_{\text{T}}$ $K^*$ due to the $p_{\text{T}}$ dependent absorption of the decay daughter hadrons. We conclude that the observed suppression of reconstructable $K^{*}$ resonances provides a strong sign of an extended hadronic rescattering stage at all investigated energies. Its duration increases from peripheral to central reactions as expected. 
Following a method, suggested by the STAR experiment, the "duration" of the hadronic stage is extracted using the $K^*/K$ ratios at chemical and kinetic freeze-out. The resulting lifetimes are in good agreement with the experimental data, but much shorter than the actual lifetime of the hadronic phase in the transport simulation. This indicates that the experimental method to estimate the life time of the hadronic stage is too simplified. 
\end{abstract}

\maketitle

\section{Introduction}
Heavy-ion collisions are today's main tool to explore the bulk properties of Quantum Chromodynamics (QCD) and strongly interacting matter at unprecedented temperatures and densities. However, the ultra-short time scales ($\approx 10^{-23}$ s) and extremely small distance scales ($\approx 10^{-15}$ m) do not allow to map out the space-time features of such collisions directly. Thus, one has to employ indirect methods to extract information about the space-time picture of the evolution. A prominently employed tool to extract information on the space-time extent of the emission source of particles, often called the region of homogeneity, is Hanbury-Brown-Twiss interferometry (HBT) \cite{HanburyBrown:1954amm,Goldhaber:1960sf,Bauer:1992ffu,Pratt:1986cc}. HBT correlation measurements of e.g. $\pi\pi$ correlations can be used to infer the volume and duration of the pion emission source. In this context HBT analyses of experimental data suggest a lifetime of the region of homogeneity of $\approx 5-15$ fm/c \cite{STAR:2014shf,PHENIX:2014dmi,Armesto:2015ioy,HADES:2019lek}. 

In contrast to these quantum correlations, one may also scrutinize short-lived hadron resonances to probe the length of the hadronic stage. Resonances like the $\rho$, $f_0$, $K^*$, $\phi$, $\Lambda^*$, etc. provide, due to their varying life times between $1-50$ fm/c, a differential measure of the extent of the hadronic stage. The main idea here is that the resonances after their creation decay and their yield is reconstructed from the daughter hadrons that have not undergone a further interaction in the hadronic medium. Thus, the ratio of the resonance to its ground state provides a space-time dependent measure of the interaction strength with the hadronic phase and thus the suppression is in a very simple picture proportional to $K^*/K \approx \exp{-\int_{t_{\text{chem}}}^{t_{\text{kin}}} \Gamma(t)~ \text{d}t} $, with $t_{\text{chem}}$ ($t_{\text{kin}}$) being the chemical (kinetic) freeze-out time, while $\Gamma(t)$ is the space-time dependent damping rate. In the most simple approximation, $\Gamma(t)$ is the vacuum decay width of the resonance\footnote{It is important to note that the simple assumption of $\Gamma(t)=\Gamma^{\text{Res}}_{\text{vac}}=$ const. is not valid, if regeneration and or collisional broadening in the medium is included. Especially for $K^*$ mesons substantial regeneration is present in a pion rich environment via the channel $K^*\rightarrow K+\pi$ followed by $K+\pi' \rightarrow K^*$, where $\pi$ and $\pi'$ are different pions.}.  This apparent suppression of reconstructable resonance states has been theoretically predicted \cite{Torrieri:2001ue,Rafelski:2001hp,Bleicher:2002dm,Markert:2002rw,Knospe:2015rja,Knospe:2015nva,Ilner:2016xqr} at various energies and has also been observed experimentally \cite{STAR:2004bgh,Markert:2005jv,STAR:2008twt,HADES:2013sfy,Knospe:2015rja,Knospe:2015nva,ALICE:2018ewo,ALICE:2022zuc,ALICE:2023ifn}. A nice advantage of vector meson resonances, e.g. $\rho^0, \phi, J/\Psi$ is that their time integrated yield could also be measured using their decay into di-leptons \cite{Rapp:1997fs,Song:1995wy,Endres:2015fna,Staudenmaier:2017vtq}, thus providing an additional handle to constrain the hadronic stage and its evolution. Examples for the investigative power of resonances beyond their suppression are e.g. their participation in the intermediate flow \cite{Reichert:2023eev} or their potential to investigate the kinetic freeze-out \cite{Reichert:2019lny,Motornenko:2019jha,Reichert:2020uxs,Reichert:2022uha} or even the structure of the resonance \cite{ALICE:2023cxn,Reichert:2024stb}. 

In this paper, we explore the suppression of $K^*(892)$ resonances in Au+Au reaction for collision energies ranging $\sqrt{s_{\text{NN}}}=3-39$ GeV. This energy range has been under investigation at the RHIC collider and its lower end will be further analysed by the future FAIR facility in the CBM experiment. 

For our study, we employ the UrQMD model (version 3.5) \cite{Bass:1998ca,Bleicher:1999xi,Bleicher:2022kcu}. All calculations were carried out using UrQMD in cascade mode and without hybrid hydrodynamic evolution. For the present study it is important that UrQMD includes all relevant resonance decays and their regeneration during the hadronic stage of the reaction. Resonances are reconstructed following the paths of the decay daughters from the space-time point of decays through the future of the simulation. Only if both decay daughters (here the $K$ and the $\pi$ from the $K^*$ decay) leave the reaction zone without further interaction, the $K^*$ is reconstructable. This method is well tested and has been applied previously \cite{Bleicher:2002dm,Knospe:2015rja,Knospe:2015nva}. The results are compared to the world data, with a special focus on the recently measured STAR data. 

\section{Energy dependence of the $(K^{*0} + \Bar{K}^{*0})/(K^+ + K^-)$ ratios}

\begin{figure} [t]
    \centering
    \includegraphics[width=\columnwidth]{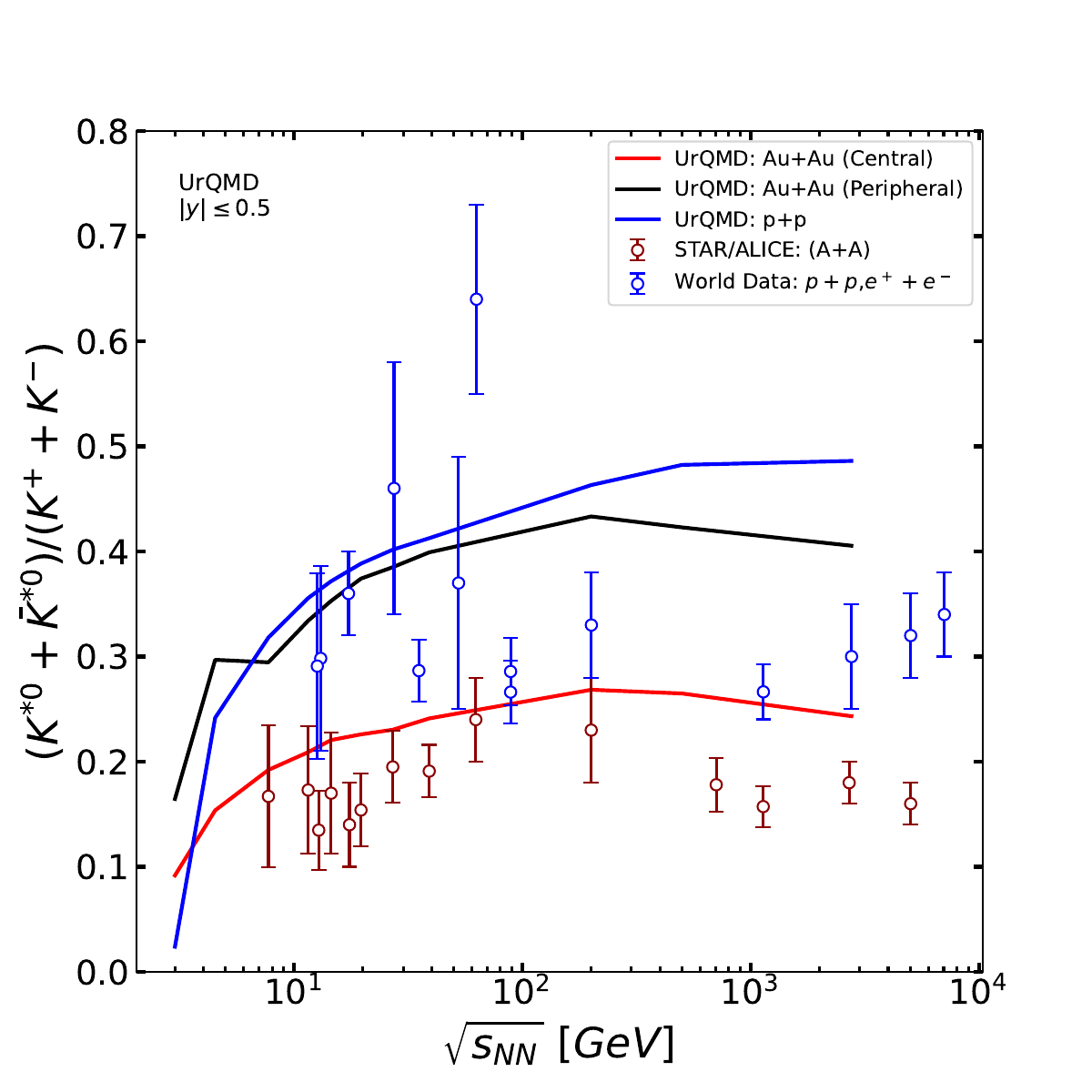}
    \caption{[Color online] Ratio of $(K^{*0} + \Bar{K}^{*0})/(K^+ + K^-)$ at midrapidity as a function the center-of-mass energy $\sqrt{s_\mathrm{NN}}$ in p+p (blue line) and Au+Au (central collisions: red line, peripheral collisions: black line) collisions from UrQMD. Experimental data (symbols) for each system are taken from: e+e \cite{ARGUS:1993ggm,Pei:1996kq,Hofmann:1988gy,SLD:1998coh}, p+p \cite{Aguilar-Benitez:1991hzq,STAR:2004bgh,AnnecyLAPP-CERN-CollegedeFrance-Dortmund-Heidelberg-Warsaw:1981whv,AxialFieldSpectrometer:1982btk}, d+Au \cite{STAR:2008twt}, p+Au \cite{ALICE:2016sak,ALICE:2021uyz} and C+C, Si+Si \cite{NA49:2011bfu}, Au+Au \cite{STAR:2004bgh,STAR:2010avo,STAR:2022sir} and Pb+Pb \cite{ALICE:2014jbq,ALICE:2017ban,ALICE:2019xyr}.}
    \label{fig:Kratio_sNN}
\end{figure}

Figure \ref{fig:Kratio_sNN} shows the ratio of $(K^{*0} + \Bar{K}^{*0})/(K^+ + K^-)$ at midrapidity as a function the center-of-mass energy $\sqrt{s_\mathrm{NN}}$ in p+p (blue line) and Au+Au (central collisions: red line, peripheral collisions: black line) collisions from UrQMD. We compare to the full breadth of available experimental data (symbols), ranging from $e^+e^-$ \cite{ARGUS:1993ggm,Pei:1996kq,Hofmann:1988gy,SLD:1998coh}, p+p \cite{Aguilar-Benitez:1991hzq,STAR:2004bgh,AnnecyLAPP-CERN-CollegedeFrance-Dortmund-Heidelberg-Warsaw:1981whv,AxialFieldSpectrometer:1982btk}, d+Au \cite{STAR:2008twt}, p+Au \cite{ALICE:2016sak,ALICE:2021uyz} to C+C, Si+Si \cite{NA49:2011bfu}, Au+Au \cite{STAR:2004bgh,STAR:2010avo,STAR:2022sir} and Pb+Pb \cite{ALICE:2014jbq,ALICE:2017ban,ALICE:2019xyr}.

As expected, the resonance ratios increase strongly with center-of-mass-energy at the lower beam energies and then level-off towards higher energies. The calculations and the experimental data are generally in line, although the uncertainties in the data are rather large. For peripheral Au+Au reactions, we observe a similar pattern as for p+p reactions, both qualitatively and quantitatively. This is in stark contrast to the results for central Au+Au reactions. Over the whole investigated energy range, (both in experiment and in the simulations) a strong suppression of the $K^*/K$ in central collisions, in comparison to p+p reactions and peripheral Au+Au collisions, is observed. We further note that the $K^*/K$ ratio exhibits a weak maximum around $\sqrt{s_{\text{NN}}}\approx 200$ GeV in peripheral and central Au+Au reactions, which is not present in the case of p+p reactions. This can be understood, because the $K^*/K$ ratio in p+p has already reached its asymptotic value, while the increase of the fireball volume leads to a suppression of the reconstructable $K^*$ resonances at very high energies.

\section{Energy and centrality dependence of the $K^*$ yields}

In the following, we focus on a differential study, that is possible due to the new STAR data. We start by a direct comparison of the $K^*$ yields for various energies and centralities as explored by the STAR experiment. The STAR experiment has recently released new data concerning the $K^{*}$ resonance, covering a range of beam energies (${\sqrt{s_{\text{NN}}} = 7.7,\ 11.5,\ 14.5,\ 19.6,\ 27\ \mathrm{ and }\ 39}$~GeV) \cite{STAR:2022sir}. We further include the energies $\sqrt{s_{\text{NN}}}$ = 3.0 and 4.5 GeV to bridge to the RHIC-BES II energies and to the planned FAIR facility. The simulations are performed for minimum bias collisions of Au+Au. To compare to the STAR data which are presented in terms of the MC Glauber number of participants ($\langle N_\text{part}\rangle$), we extract $N_\text{part}$ based on the Glauber model for each event as done in the experimental data\footnote{We want to point out the well known fact, that the number of participants extracted using the Glauber model is not identical to the number of participants obtained from momentum space cuts or by summing all nucleons that have interacted.}.

\begin{figure} [t]
    \centering
    \includegraphics[width=\columnwidth]{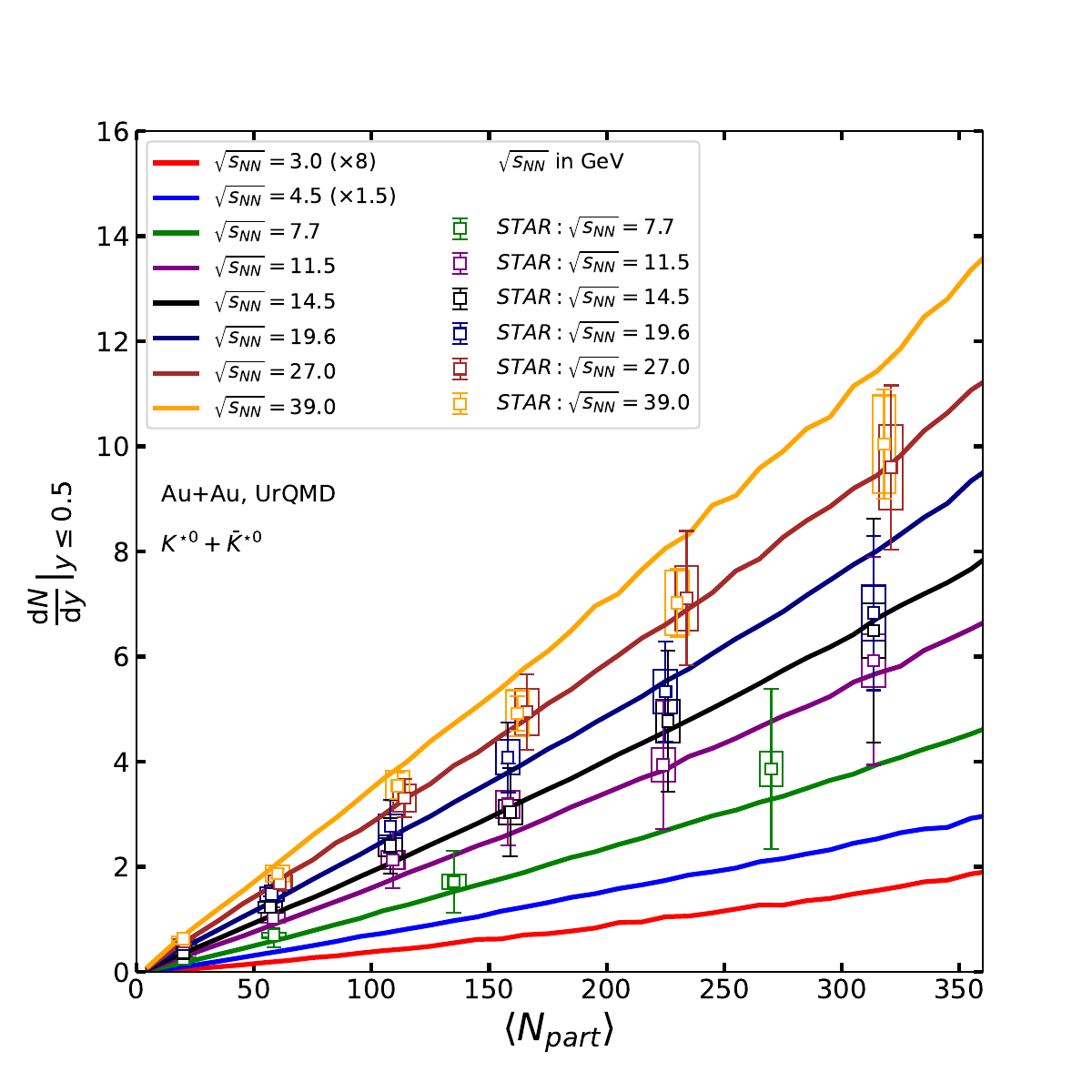}
    \caption{[Color online] Centrality dependence of the midrapidity yields $\text{d}N/\text{d}y|_{|y|\leq0.5}$ of $K^{*0} + \Bar{K}^{*0}$ as a function of $N_{\text{part}}$ at $\sqrt{s_{\text{NN}}}$ = 3.0, 4.5, 7.7, 11.5, 14.5, 19.6, 27 and 39 GeV in Au+Au collisions. UrQMD results are shown as lines, experimental data from \cite{STAR:2022sir} are shown as open squares with error bars. Note that the UrQMD results for the lowest energies are scaled up by factors $8$ and $1.5$ for better visibility.}
    \label{fig:Kyield_Npart}
\end{figure}

We begin with a comparison of the midrapidity yields ($\text{d}N/\text{d}y|_{|y|\leq0.5}$) of the $K^{*0} + \Bar{K}^{*0}$ shown in Figure \ref{fig:Kyield_Npart}. The lines show UrQMD calculations, while the open symbols show the experimental data from STAR \cite{STAR:2022sir} as open squares (with statistical and systematic error bars) as a function of centrality. As expected, the yields increase with collision energy and towards more central reactions. The overall comparison between model simulation and experimental data is very good within the error bars.

\begin{figure} [t]
    \centering
    \includegraphics[width=\columnwidth]{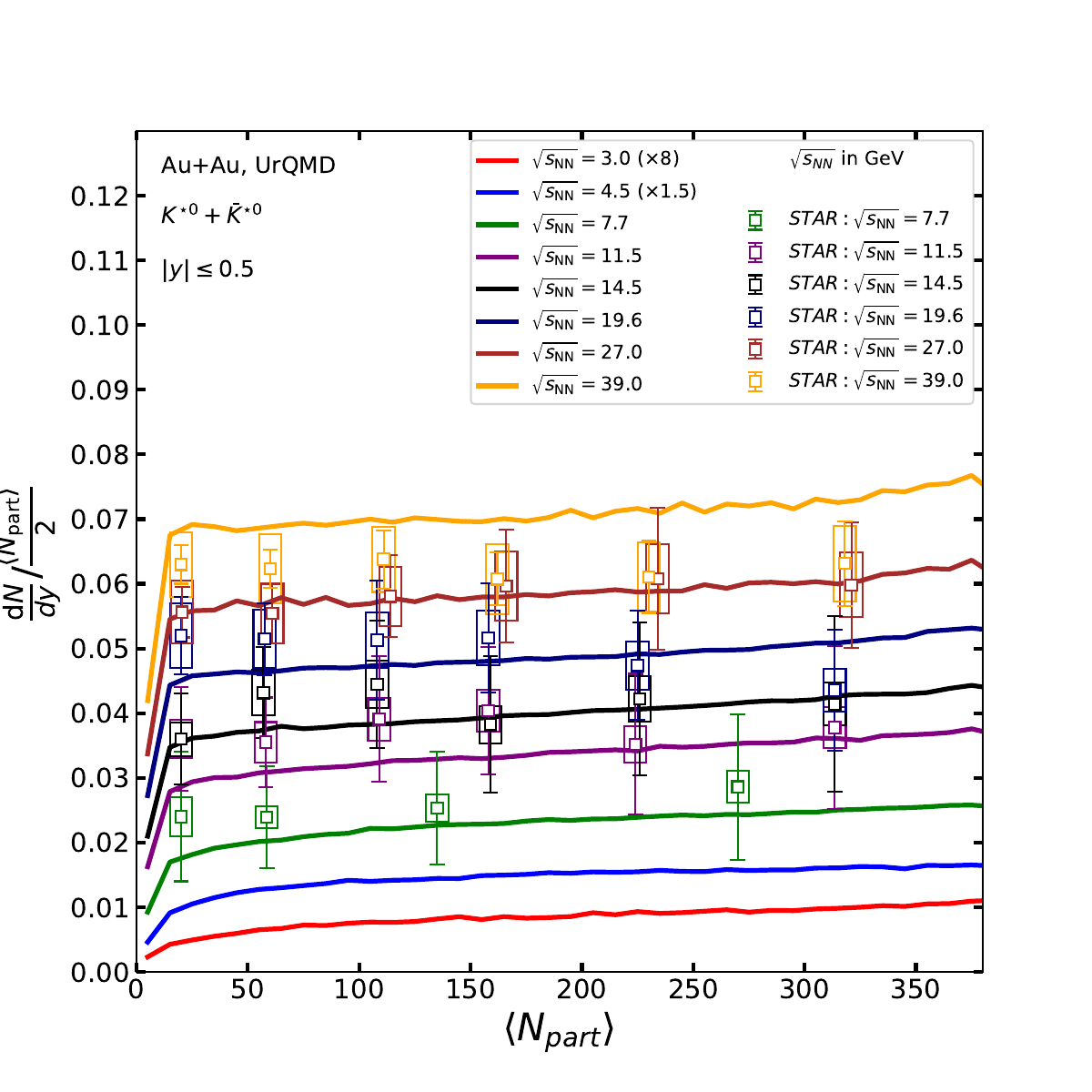}
    \caption{[Color online] Centrality dependence of the midrapidity yields scaled by the number of participant pairs $\text{d}N/\text{d}y|_{|y|\leq0.5}/(\langle N_\text{part}\rangle/2)$ of $K^{*0} + \Bar{K}^{*0}$ as a function of $N_{\text{part}}$ at $\sqrt{s_{\text{NN}}}$ = 3.0, 4.5, 7.7, 11.5, 14.5, 19.6, 27 and 39 GeV in Au+Au collisions. UrQMD results are shown as lines, experimental data from \cite{STAR:2022sir} are shown as open squares with error bars. Note that the UrQMD results for the lowest energies are scaled up by factors $8$ and $1.5$ for better visibility.}
    \label{fig:Kyield_Npart_scaled}
\end{figure}

Next we turn to the centrality scaling of the $K^{*0} + \Bar{K}^{*0}$ yields. To this aim Fig.~\ref{fig:Kyield_Npart_scaled} shows the yields scaled by the number of participant nucleon pairs ($\text{d}N/\text{d}y|_{|y|\leq0.5}/(\langle N_\text{part}\rangle/2)$) for various collision energies. Again the lines show UrQMD calculations, while the open symbols show the experimental data from STAR \cite{STAR:2022sir} (with statistical and systematic error bars) as a function of centrality. One observes that the scaling of the $K^*$ yields with centrality is also well captured by the transport simulation.

\begin{figure} [t]
    \centering
    \includegraphics[width=\columnwidth]{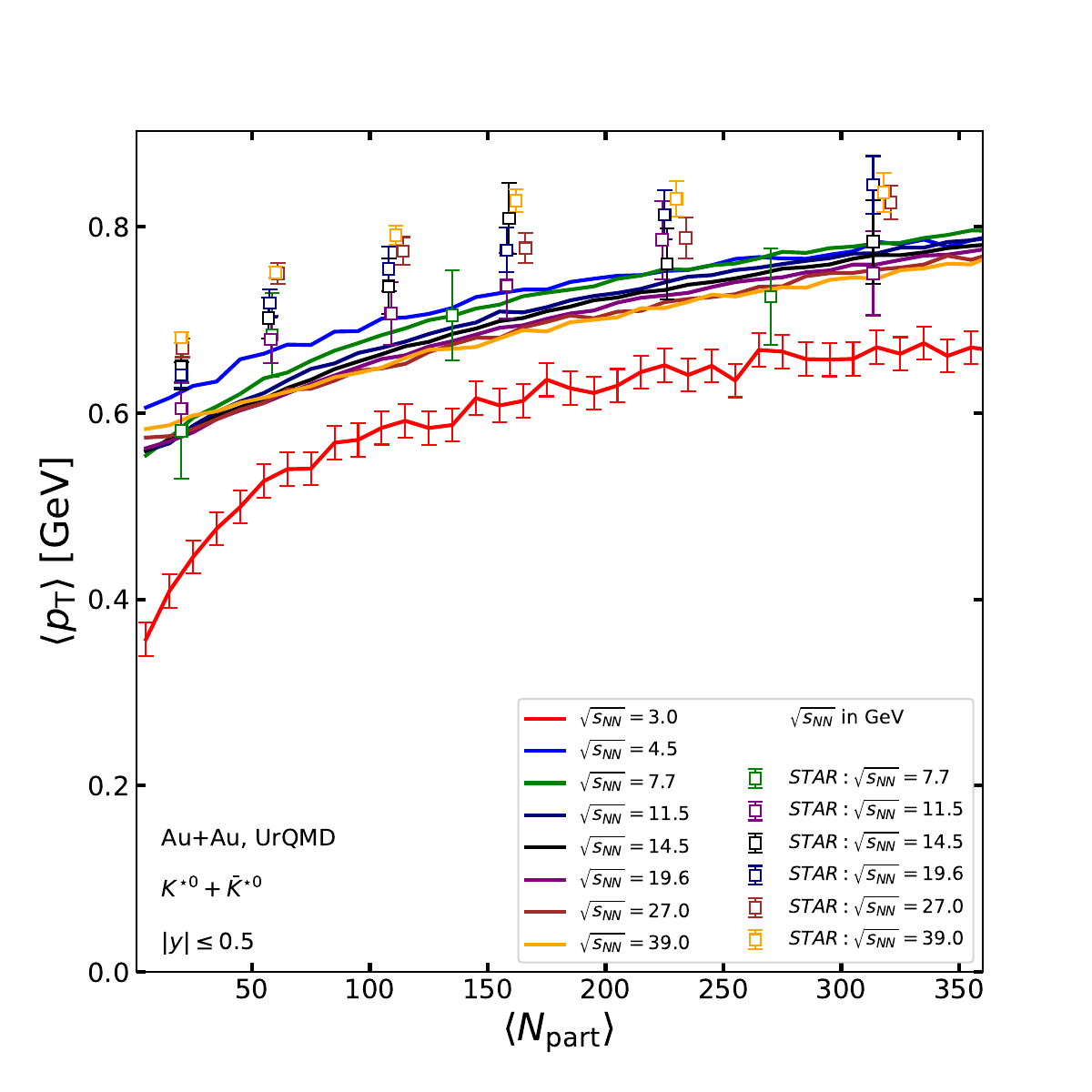}
    \caption{[Color online]  Centrality dependence of the mean transverse momenta $\langle p_\text{T} \rangle|_{|y|\leq0.5}$ at midrapidity of $(K^{*0} + \Bar{K}^{*0})$ as a function of $N_{\text{part}}$ at $\sqrt{s_{\text{NN}}}$ = 3.0, 4.5, 7.7, 11.5, 14.5, 19.6, 27 and 39 GeV in Au+Au collisions. UrQMD results are shown as lines, experimental data from \cite{STAR:2022sir} are shown as open squares with error bars.}
    \label{fig:Kptrans_Npart}
\end{figure}

Finally, Fig.~\ref{fig:Kptrans_Npart} shows the mean transverse momenta $\langle p_\text{T}\rangle$ of the $K^*$ as a function of centrality for the range of energies under study. The energy dependence of the mean transverse momentum is surprisingly weak (except for the calculation at $\sqrt{s_{\text{NN}}} = 3.0$ GeV). For all energies one clearly observes an increase of the mean transverse momenta towards central reactions. This observation is in line with the findings at different energies previously studied. It is usually attributed to the effect that the transverse momentum spectra are apparently 'hardened' due to the higher absorption probability of low $p_\text{T}$ decay products for more central collisions \cite{Bleicher:2002dm,Knospe:2015nva,Reichert:2019lny,Knospe:2021jgt,Reichert:2022uha}. 

\section{Resonance to ground state ratios}

\begin{figure} [t]
    \centering
    \includegraphics[width=\columnwidth]{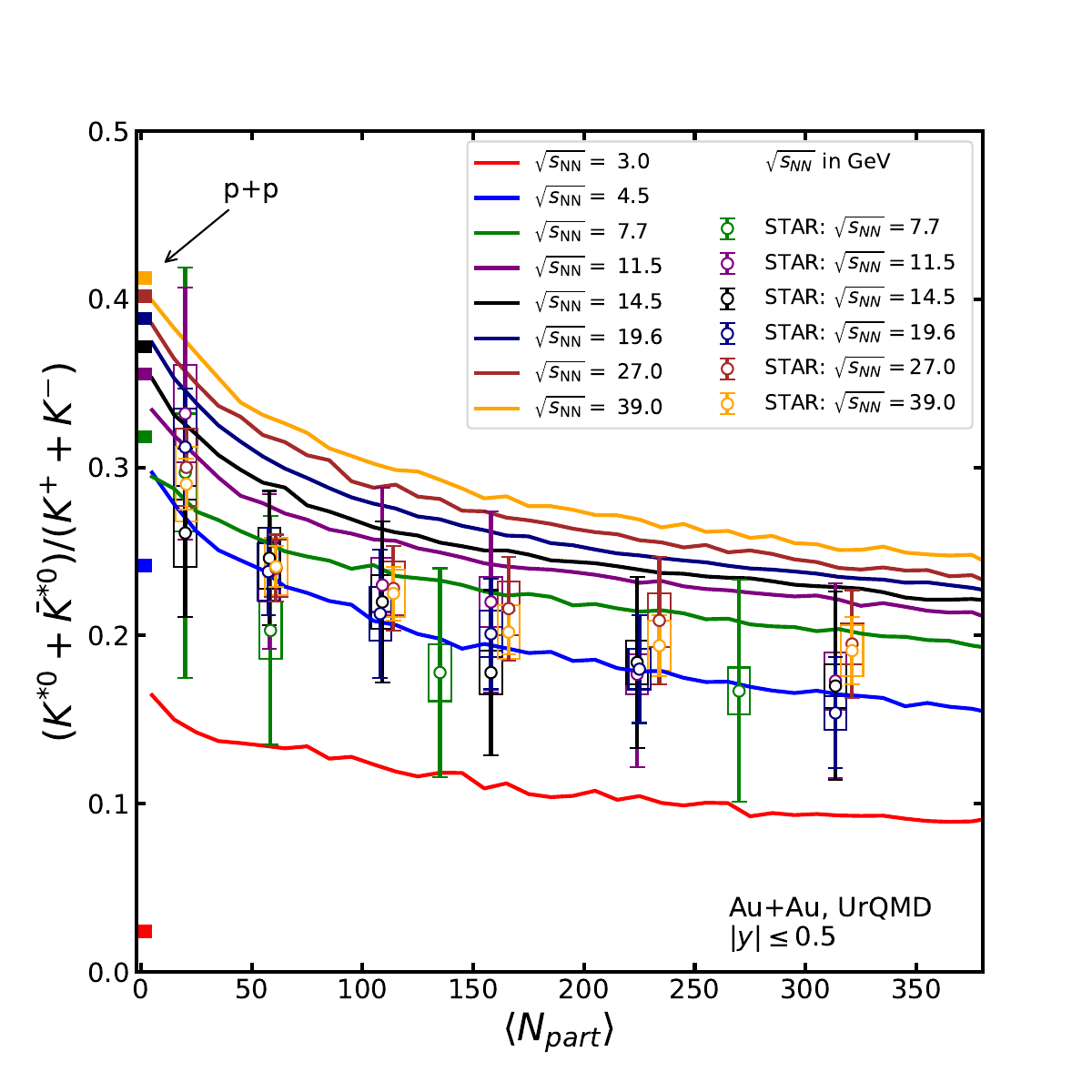}
    \caption{[Color online] Ratio of $(K^{*0} + \Bar{K}^{*0})/(K^+ + K^-)$ at midrapidity as a function of $N_{\text{part}}$ at $\sqrt{s_{\text{NN}}}$ = 3.0, 4.5, 7.7, 11.5, 14.5, 19.6, 27 and 39 GeV in p+p (filled squares) and Au+Au (lines) collisions from UrQMD. Experimental data from \cite{STAR:2022sir} are shown as open circles with error bars.}
    \label{fig:Kratio_Npart}
\end{figure}

After establishing agreement of the mid-rapidity yields and mean transverse momenta of the $K^*$ resonances with the experimental data, we can now move on and calculate and compare resonances ratios, i.e. the ratio $(K^{*0}+\Bar{K}^{*0})/(K^+ + K^-)$.
Fig.~\ref{fig:Kratio_Npart} shows the ratio of $(K^{*0} + \Bar{K}^{*0})/(K^+ + K^-)$ as a function of $N_{\text{part}}$ at ${\sqrt{s_{\text{NN}}} = 3.0,\ 4.5,\ 7.7,\ 11.5,\ 14.5,\ 19.6,\ 27\ \mathrm{and}\ 39}$ GeV in p+p (filled squares) and Au+Au (lines) collisions from UrQMD. Experimental data from \cite{STAR:2022sir} are shown as open circles with error bars. 
As expected, the $(K^{*0} + \Bar{K}^{*0})/(K^+ + K^-)$ ratio shows a decrease towards central collisions by 30\% at all investigated energies. This indicates that the rescattering effect on the daughter particles leads to a signal loss for the $K^*$ reconstruction in line with the previously observed increase in mean $p_{\text{T}}$.  It is noteworthy that only at higher energies ($\sqrt{s_{\text{NN}}}\geq 7.7$ GeV) the ratio in very peripheral Au+Au collisions ($N_\text{part}\approx 25$) tends towards the value from p+p collisions. At lower energies the p+p value of the ratio is smaller than in peripheral Au+Au reactions suggesting that threshold effects start to kick in (note that the threshold for $K^*$ production is $m_{K^*}+m_\Lambda+m_p=2.945$ GeV and Fermi-momenta are absent in p+p collisions). The $K^*/K$ ratio thus exhibits a non-monotonic centrality dependence  at the lower collision energies. The ratios compare rather well with the ratios measured by STAR \cite{STAR:2022sir} within the reported error bars. However, one should note that towards the higher collision energies, the simulation yields slightly less suppression for central and peripheral reactions than observed in the experimental data. Since the $K^*$ yields in peripheral and central reactions where in line with the data (see Figs. \ref{fig:Kyield_Npart} and \ref{fig:Kyield_Npart_scaled}) the overestimation of the $K^*/K$ can be traced back to the known under prediction of the charged Kaon yields \cite{Petersen:2008kb}.

Let us now come to the interpretation of the suppression of the observed $K^*$ resonances in central reactions. Within the UrQMD model, the suppression is clearly linked to the signal loss that emerges from the rescattering of the daughter particles. This interpretation is supported and in line with with previous model studies, see e.g. \cite{Bleicher:2002dm,Knospe:2015nva,Ilner:2016xqr,Werner:2018yad,Knospe:2021jgt}. Also the simple test of switching-off hadronic rescattering further supports rescattering of the daughter hadrons as the dominant process for the suppression (loss of the signal of reconstructable resonances). However, also alternative scenarios have been put forward that do not assume a signal loss due to rescattering, but attribute the suppression to the cooling of the hadronic system in partial chemical equilibrium which suppress the $K^*$ via a substantially lowered (chemical) freeze-out temperature, see e.g. \cite{Motornenko:2019jha,LeRoux:2021adw}.

\section{Estimates for the duration of the hadronic stage}
In this final section we will discuss the possibility and problems to extract the lifetime of the hadronic stage. 

The resonance's time evolution can be described by a kinetic equation: 
\begin{equation}
    N^{K^*} (t)=N^{K^*}_\text{chem}-\Gamma^{K^*} N^{K^*} \text{d}t - L_\text{coll} ~\text{d}t + G_\text{reg} ~\text{d}t,
\end{equation}
where $\text{d}t$ is the time interval to be integrated from the chemical freeze-out to the kinetic freeze-out (life time of the hadronic phase, $\Delta t_\text{hadronic}$). $\Gamma^{K^*}$ is the width of the $K^*$, $ L_\text{coll}$ is the loss rate due to collisions of the $K^*$ and $G_\text{reg}$ is a gain rate, e.g. $K+\pi \rightarrow K^*$. 
In addition the decaying $K^*\rightarrow h_1 h_2$ needs to be observable, which can be captured by a scattering probability of the decays daughters $P(h_i,\rho_\text{hadron},\Delta_\text{escape})$, see e.g. \cite{Torrieri:2001ue,Torrieri:2001tg}. 

Assuming that that the interaction probability for the decay daughters $P(h_i,\rho_\text{hadron},\Delta_\text{escape})$ is large and the gain  and loss  rates can be neglected during the hadronic stage one obtains the relation 
\begin{equation}
N^{K^*}(t_\text{kin})/N^{K^*}(t_\text{chem})=\exp(-\Gamma^{K^*}\Delta t_{\text{hadronic}}).
\end{equation}
With the further assumption that $\Gamma^{K^*}$ in the medium can be approximated by the vacuum width, this approximation was used by STAR \cite{STAR:2002npn,STAR:2022sir} to obtain a lower bound for the duration of the hadronic rescattering phase in heavy-ion collisions. As only final state particles are measurable, the yields of resonances at the chemical decoupling surface cannot be measured directly. To overcome this issue, STAR has further assumed that the ratio at the chemical freeze-out in a nucleus+nucleus collision is close to the ratio measured in $pp$ and/or $e^+e^-$ reactions and used an energy independent constant fit as the $K^*/K$ ratio baseline at chemical freeze-out. Following this simple approach leads to the relation
\begin{align}
    \left( \frac{K^*}{K} \right)\bigg|_{\text{KFO}} &= \left( \frac{K^*}{K} \right)\bigg|_{\text{CFO}} \times \exp\left( - \frac{\Delta t_{\text{hadronic}}}{\tau_{K^*}} \right).
\end{align}
in which the short-hand notation $K^*/K$ represents the $(K^{*0}+\Bar{K}^{*0})/(K^+ + K^-)$ ratio, $\Delta t_{\text{hadronic}}$ is the lifetime of the hadronic rescattering phase, $\tau_{K^*} = 1/\Gamma^{K^{\star}}_{\text{vac}}$ is the (vacuum) lifetime of the $K^*$ resonance and the labels ``CFO" and ``KFO" refer to the chemical and kinetic freeze-out, respectively. This procedure is then applied to the calculated centrality dependent $K^*/K$ ratios to obtain the estimates of the time between chemical and kinetic decoupling. 

\begin{figure} [t]
    \centering
    \includegraphics[width=\columnwidth]{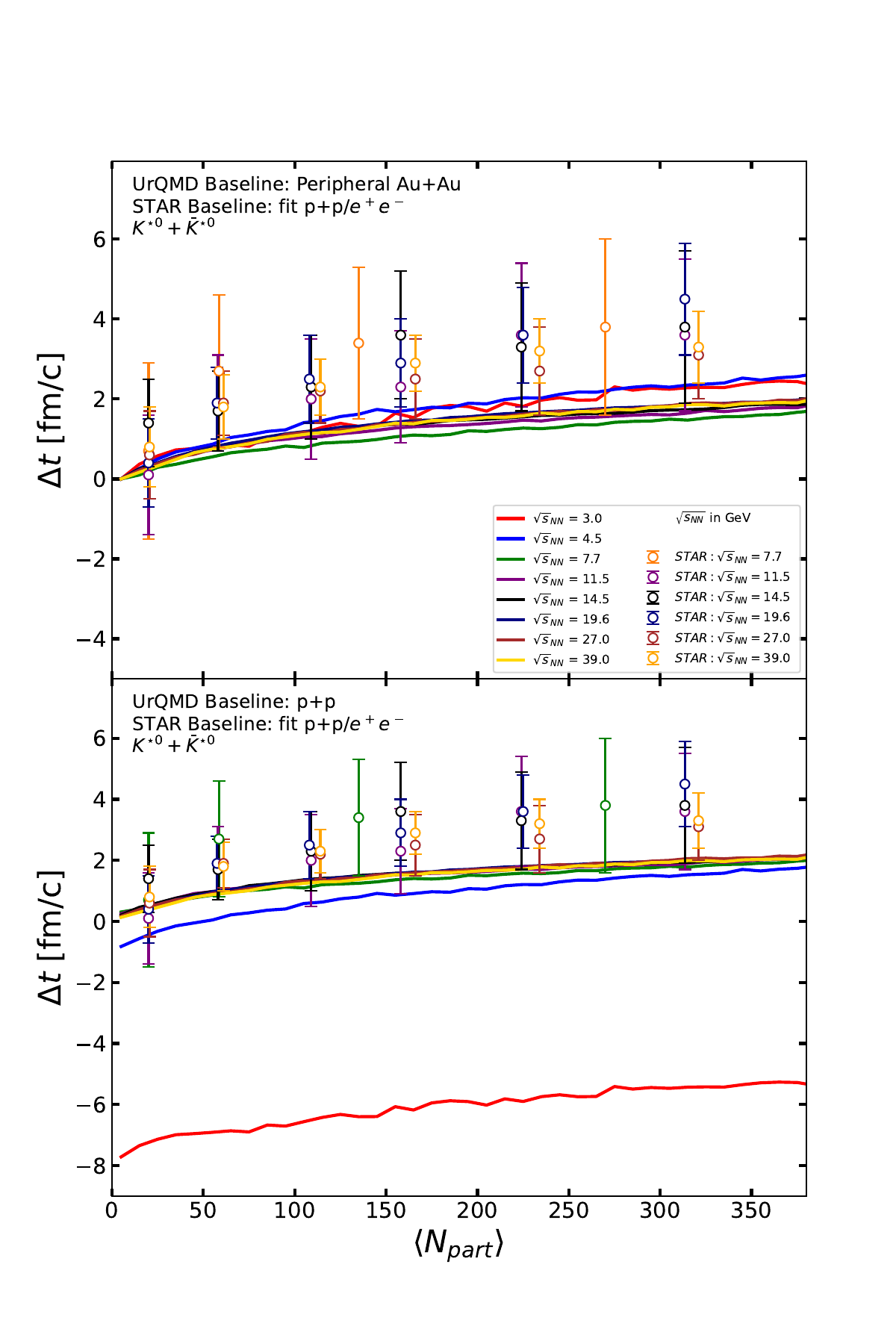}
    \caption{[Color online] Lower estimate for the duration of the hadronic rescattering phase $\Delta t$ as a function of $N_{\text{part}}$ at $\sqrt{s_{\text{NN}}}$ = 3.0, 4.5, 7.7, 11.5, 14.5, 19.6, 27, 39 and 62.4 GeV in Au+Au (lines) collisions from UrQMD. Experimental data from STAR \cite{STAR:2004bgh,STAR:2010avo,STAR:2022sir} and ALICE \cite{ALICE:2014jbq,ALICE:2017ban,ALICE:2019xyr} are shown as open circles. Statistical and systematical error are combined. The baseline $K^*/K$ ratio is extracted from an energy independent fit to the available $p+p$ and $e^+e^-$ data. The top and bottom Figures compare different assumptions for the chemical freeze-out baseline. Top: The chemical freeze out $K^*/K$ ratios are assumed to be given by the most peripheral Au+Au reaction, bottom: The chemical freeze out $K^*/K$ ratios are assumed to be given by the p+p reactions.} 
    \label{fig:deltat_Npart}
\end{figure}
\begin{figure} [t]
    \centering
    \includegraphics[width=\columnwidth]{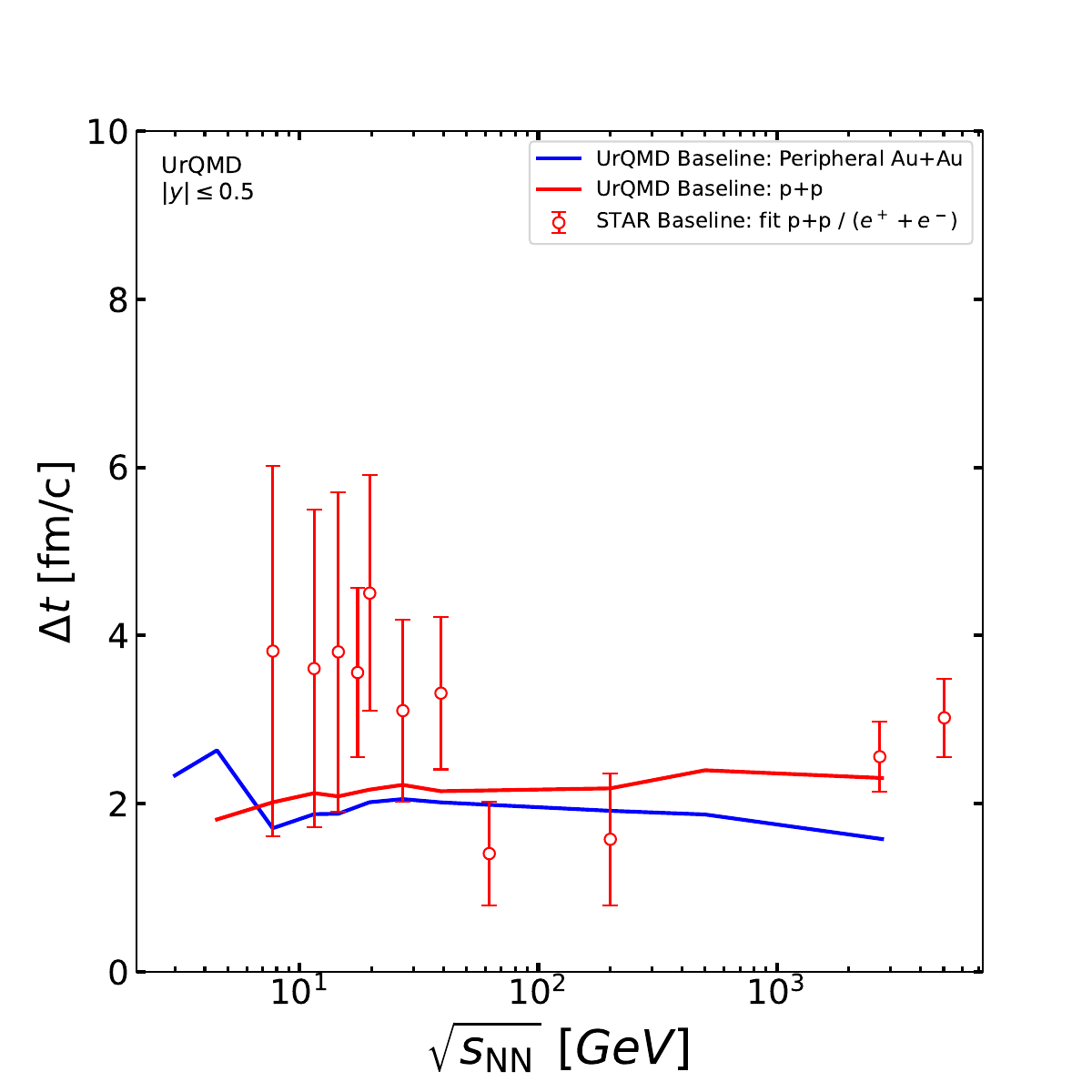}
    \caption{[Color online] Lower estimate for the duration of the hadronic rescattering phase $\Delta t$ as a function of the center-of-mass energy $\sqrt{s_\mathrm{NN}}$ in Au+Au (lines) collisions from UrQMD. Experimental data for the lifetime estimate based on the $K^*/K$ ratio are taken from \cite{ARGUS:1993ggm,Pei:1996kq,Hofmann:1988gy,SLD:1998coh,Aguilar-Benitez:1991hzq,STAR:2004bgh,AnnecyLAPP-CERN-CollegedeFrance-Dortmund-Heidelberg-Warsaw:1981whv,AxialFieldSpectrometer:1982btk,STAR:2008twt,ALICE:2016sak,ALICE:2021uyz,NA49:2011bfu,STAR:2004bgh,STAR:2010avo,STAR:2022sir,ALICE:2014jbq,ALICE:2017ban,ALICE:2019xyr} and shown as open circles. The red and blue lines compare UrQMD calculations using different baselines for the $K^*/K$ ratio at chemical freeze-out (see previous Figure).}
    \label{fig:deltat_sNN}
\end{figure}
Before presenting the results let us address some problems with this method:
\begin{enumerate}
    \item The assumptions leading to the final equation used by STAR are rather strong and are not supported by a microscopic analysis of the simulated heavy ion collision. In our calculations the gain and loss terms in the hadronic phase are substantial due to the interactions of the Kaons with surrounding pions, this also leads to a sizeable regeneration rate for $K^*$. These limitations can be overcome in the kinetic equation approach by using a full network of rate equations as it is e.g. done for light nuclei, and resonance ratios at the LHC \cite{Motornenko:2019jha,Neidig:2021bal,Lokos:2022jze}.
    \item The method does not account for relativistic expansion of the fireball, leading to a relative velocity of the $K^*$ with respect to the center of the fireball. A corresponding $\gamma$ factor has been adopted e.g. in Ref. \cite{ALICE:2019xyr}. However, in order to keep consistency with the STAR data \cite{STAR:2022sir}, this additional $\gamma$ factor will not be considered here.
    \item Due to the lack of high quality $K^*$ data in elementary reactions, the STAR collaboration uses a constant fit through the available $pp$ and $e^+e^-$ data as an energy independent baseline of the ratio at chemical freeze-out. This assumption is certainly not justified when the collision energy decreases towards the threshold and the $K^*/K$ ratio in p+p reactions is much smaller than in peripheral Au+Au reactions (see Fig. \ref{fig:Kratio_sNN}).
    \item It is usually argued that the time estimate based on this approach yields a lower boundary for the duration of the hadronic rescattering phase. While this is true, the restriction is even more severe and triggers only on the last generation of resonance, thus the 'view into the past' of the fireball is limited to the lifetime of the last generation of resonances.  
\end{enumerate}

After these disclaimers, Fig.~\ref{fig:deltat_Npart} shows the lower estimate for the duration of the hadronic rescattering phase $\Delta t_{\text{hadronic}}$ as a function of $N_{\text{part}}$ at ${\sqrt{s_{\text{NN}}} = 3.0,\ 4.5,\ 7.7,\ 11.5,\ 14.5,\ 19.6,\ 27\ \textrm{and}\ 39}$~GeV in central Au+Au  collisions from UrQMD (lines) in comparison to the experimental data from STAR \cite{STAR:2022sir} that are shown as open circles with error bars (assuming the energy independent resonance ratio at chemical freeze-out discussed above). In UrQMD we have used two different estimates for the $K^*/K$ ratio at chemical freeze-out: I) In Fig.~\ref{fig:deltat_Npart} (top) we present $\Delta t_{\text{hadronic}}$ calculated under the assumption that the resonance ratio at chemical decoupling equals the peripheral Au+Au value, while II) in Fig.~\ref{fig:deltat_Npart} (bottom) we depict the hadronic life time under the assumption that the $K^*/K$ ratio at chemical freeze-out equals the value obtained in p+p collisions. 

The time estimates for the entire hadronic rescattering phase reveal a clear pattern. For most peripheral collisions, the lifetime is nearly zero due to the construction of the equation. As collisions become more central, the lifetime increases, with values increasing to 2 fm/c (similar to the life time of the $K^*$) for all energies (essentially independent of the baseline used to obtain the ratio at chemical freeze-out). Note, however, that the 'lifetime' extracted at the lowest energies of 3 GeV and 4.5 GeV becomes negative, if one assumes that the p+p reactions provide the baseline value for the chemical freeze-out reference (Fig. \ref{fig:deltat_Npart} (bottom)). As discussed above this is due to the fact that the elementary reaction does not allow for sufficient $K^*$ production such close to the threshold. This reinforces the limitations of the current experimental approach.

Let us finally analyze the hadronic 'lifetime' as a function of collision energy. Here Fig.~\ref{fig:deltat_sNN} summarizes the (lower) estimate for the duration of the hadronic rescattering phase $\Delta t_{\text{hadronic}}$ as a function of the center-of-mass energy $\sqrt{s_\mathrm{NN}}$ in Au+Au (the red line denotes the p+p baseline, the blue line denotes the peripheral Au+Au baseline) collisions from UrQMD. Experimental data for the lifetime estimate based on the $K^*/K$ ratio are taken from \cite{ARGUS:1993ggm,Pei:1996kq,Hofmann:1988gy,SLD:1998coh,Aguilar-Benitez:1991hzq,STAR:2004bgh,AnnecyLAPP-CERN-CollegedeFrance-Dortmund-Heidelberg-Warsaw:1981whv,AxialFieldSpectrometer:1982btk,STAR:2008twt,ALICE:2016sak,ALICE:2021uyz,NA49:2011bfu,STAR:2004bgh,STAR:2010avo,STAR:2022sir,ALICE:2014jbq,ALICE:2017ban,ALICE:2019xyr} and are shown as open circles.

The estimate of the duration of the hadronic rescattering phase in central Au+Au collisions from the UrQMD model  follows the trend seen in the STAR data within error bars when employing the same estimation method. Let us note however, that the duration of the hadronic expansion stage from chemical to kinetic freeze-out in the model simulations is substantially longer than these estimates \cite{Knospe:2015nva,Reichert:2020yhx,Reichert:2022qvt} indicating the limitations of the experimentally chosen approach.

\section{Conclusion}
The Ultra-relativistic Quantum Molecular Dynamics model was used to simulate Au+Au and p+p collisions in the RHIC-STAR-BES and FAIR-CBM energy range. Reconstructable $K^*$ resonances where extracted and compared to the recent STAR data. The model calculations provide a good description of the yields and mean transverse momenta for the $K^*$ resonances lining up with the experimental data. The decrease of the ratio $(K^{*0}+\Bar{K}^{*0})/(K^+ + K^-)$ with centrality is well captured by the UrQMD model and attribute to the signal loss of the daughter hadrons. We have further argued that the the $K^*/K$ suppression due to the (mostly low $p_{\text{T}}$) signal loss of the daughter hadrons is in line with the increase of $\langle p_{\text{T}}\rangle$ as a function of centrality. We have further employed a simplistic approach to estimate the lower bound for the duration of the hadronic rescattering phase from chemical to kinetic freeze-out by assuming that resonance yields only decrease with an exponential decay law. The $\Delta t_{\text{hadronic}}$ estimates from the UrQMD model with this approach follow the estimates for the hadronic life time extracted from the experimental data for $K^*/K$. One should note however, that the true hadronic life time from chemical to kinetic freeze-out in the model simulation is substantially longer than these estimates.

\section*{Acknowledgments}
The authors thank A. Knospe and C. Markert for fruitful discussion during the Strangeness in Quark Matter conference 2024 in Strasbourg.
L.E. acknowledges support from the Hermann-Willkomm-Stiftung (project number 30120106). 
T.R. acknowledges support through the Main-Campus-Doctus fellowship provided by the Stiftung Polytechnische Gesellschaft Frankfurt am Main (SPTG). 
T.R. and J.S. thank the Samson AG for their support.
This article is part of a project that has received funding from the European Union’s Horizon 2020 research and innovation programme under grant agreement STRONG – 2020 - No 824093.
The computational resources for this project were provided by the Center for Scientific Computing of the GU Frankfurt and the Goethe HLR.

\end{document}